\documentclass[useAMS,usenatbib]{mn2e}
\usepackage{graphicx}
\usepackage{epsfig}
\usepackage{color}

\title[{\it Suzaku} Results of G306.3$-$0.9 \& the Gamma Ray Neighborhood]{{\it Suzaku} Analysis of the Supernova Remnant G306.3$-$0.9 and the Gamma-ray View of Its Neighborhood}
\author[A.Sezer, T.Ergin and R.Yamazaki]{A.~Sezer,$^{1}$\thanks{E-mail: aytap.sezer@avrasya.edu.tr (AS); ergin.tulun@gmail.com (TE); ryo@phys.aoyama.ac.jp (RY)}
T.~Ergin$^{2}$ and R.~Yamazaki$^{3}$\\
$^{1}$Avrasya University, Faculty of Engineering and Architecture, Department of Electrical-Electronics Engineering, 61250, Trabzon, Turkey\\
$^{2}$TUBITAK Space Technologies Research Institute, ODTU Campus, 06800, Ankara, Turkey\\
$^{3}$Department of Physics and Mathematics, Aoyama Gakuin University, 5-10-1 Fuchinobe, Sagamihara 252-5258, Japan\\
}

\begin{document}
\date{}
\pagerange{\pageref{firstpage}--\pageref{lastpage}} \pubyear{2016}
\maketitle
\label{firstpage}

\begin{abstract}
We present an investigation of the supernova remnant (SNR) G306.3$-$0.9 using archival multi-wavelength data. The {\it Suzaku} spectra are well described by  two-component thermal plasma models: The soft component is in ionization equilibrium and has a temperature $\sim$0.59 keV, while the hard component has temperature $\sim$3.2 keV and ionization time-scale $\sim$$2.6\times10^{10}$ cm$^{-3}$ s. We clearly detected Fe K-shell line at energy of $\sim$6.5 keV from this remnant. The overabundances of Si, S, Ar, Ca, and Fe confirm that the X-ray emission has an ejecta origin. The centroid energy of the Fe-K line supports that G306.3$-$0.9 is a remnant of a Type Ia supernova (SN) rather than a core-collapse SN. The GeV gamma-ray emission from G306.3$-$0.9 and its surrounding were analyzed using about 6 years of {\it Fermi} data. We report about the non-detection of G306.3$-$0.9 and the detection of a new extended gamma-ray source in the south-west of G306.3$-$0.9 with a significance of $\sim$13$\sigma$. We discuss several scenarios for these results with the help of data from other wavebands to understand the SNR and its neighborhood. 
\end{abstract}


\section{Introduction}
G306.3$-$0.9 is a new Galactic supernova remnant (SNR) that was detected at mid-IR wavelengths by {\it Spitzer} at 24 $\micron$ \citep{Ca09} and {\it WISE} at 22 $\micron$ \citep {Wr10}. \citet {Re13} presented a study of {\it Chandra}, 24 $\micron$ {\it Spitzer} and the 5 GHz Australia Telescope Compact Array (ATCA) data. The {\it Chandra} spectrum was described by a thermal plasma model (VNEI, VPSHOCK or Sedov) with the absorption column density $N_{\rm H}$ $\sim$ (1.94$-$1.96)$\times10^{22}$ cm$^{-2}$. 
In their spectral fitting, the ionization time-scale shows that the plasma is approaching ionization equilibrium, $\tau$ in the range of $\sim$(0.89$-$2.5)$\times10^{12}$ s cm$^{-3}$.

Recently, \citet {Co16} studied the northeast, central, and southwest part of the G306.3$-$0.9 using {\it XMM-Newton} and {\it Chandra} data. They found that the X-ray spectra were well represented by two absorbed VAPEC and VNEI thermal plasma models with the absorption column density $N_{\rm H}$ $\sim$ (1.40$-$1.57)$\times10^{22}$ cm$^{-2}$. They also found enhanced abundances of Si, S, Ar, Ca, and Fe in the VNEI component, indicating that the X-ray emission has an ejecta origin. The other thermal component is associated with the swept-up interstellar medium (ISM).

Using the  high-resolution {\it Chandra} data, \citet{Re13} and \citet{Co16} concluded that G306.3$-$0.9 has a semi-circular and asymmetric X-ray morphology. The X-ray emission of the southern region shows semi-circular brightened structure, while the northern emission is very weak. The image of the southern region is also consistent with radio and IR observation. \citet {Re13} and \citet {Co16} discussed the progenitor of this remnant and favored a Type Ia supernova (SN) rather than a core-collapse (CC) SN.

In the 1st SNR Catalog of the Large Area Telescope (LAT) on board {\it Fermi} Gamma Ray Space Telescope \citep{Ac16}, G306.3$-$0.9 was mentioned in the `SNRs Not Detected by the LAT' table (Table 3). However, another nearby extended source,  `G306.3$-$00.8', was reported among the `Other Detected Sources', which are not classified as SNRs. It was detected with a significance of $\sim$9$\sigma$ and the best-fitting location was found as R.A.(J2000) = 199$^{\circ}\!\!$.33 $\pm$ 0$^{\circ}\!\!$.07$^{\rm stat}$  $\pm$ 0$^{\circ}\!\!$.17$^{\rm sys}$ and decl.(J2000) = $-$62$^{\circ}\!\!$.96 $\pm$ 0$^{\circ}\!\!$.07$^{\rm stat}$  $\pm$ 0$^{\circ}\!\!$.21$^{\rm sys}$. The extension (radius of a disk) of this source was measured as 0$^{\circ}\!\!$.53 $\pm$ 0$^{\circ}\!\!$.07$^{\rm stat}$ $\pm$ 0$^{\circ}\!\!$.07$^{\rm sys}$. The spectrum was fit to a power-law (PL) that yielded a photon flux and a photon index of (7.77 $\pm$ 0.94) $\times$ 10$^{-9}$ ph cm$^{-2}$ s$^{-1}$ and $\Gamma$ = 2.56 $\pm$ 0$^{\circ}\!\!$.18$^{\rm stat}$ $\pm$ 0$^{\circ}\!\!$.1$^{\rm sys}$, respectively. When \citet{Ac16} tested two alternative IEMs (the interstellar emission model, which accounts for gamma rays produced by cosmic ray (CR) interactions with interstellar gas and radiation fields in the Milky Way) for `G306.3$-$00.8', they found out that the source had an extended gamma-ray morphology for one of the tested IEMs, and was a point-like source (was not extended) for the other IEM. 

The nearby gamma-ray source 3FGL J1317.6$-$6315 is a pulsar candidate classified by two different statistical pulsar classification methods (i.e. BLR: Boosted Logistic Regression \& RF: Random Forest) in \citet{Saz16}. The location of the source given in the 3rd Fermi Source Catalog (3FGL) is R.A.(J2000) = 199$^{\circ}\!\!$.403 and decl.(J2000) = $-$63$^{\circ}\!\!$.259 and its detection significance is about 13.5$\sigma$ \citep{Ac15}. Another significant gamma-ray source close to G306.3$-$0.9 is the HESS unidentified object, HESS J1303$-$631 \citep{Ah05, HE11}, which is interpreted as physically related to the pulsar PSR J1301$-$6305. It is assumed that the electrons radiatively cool as they propagate away from the pulsar, such that the highest-energy gamma rays are found close to the pulsar and the lower-energy particles are mainly in the extended nebula. X-ray observations revealed a pulsar wind nebula (PWN) around PSR J1301$-$6305 extending asymmetrically roughly towards the gamma-ray source. The most recent sensitivity profile derived from the H.E.S.S. Galactic Plane Survey \citep{Do16} shows that the sensitivity ranges between 1 and 2 per cent of the Crab Nebula, which corresponds to about $\sim$1$\times$10$^{-11}$ erg s$^{-1}$ cm$^{-2}$ in 1$-$10 TeV, but the point-like source significance map shows no TeV gamma-ray excess at or around the location of G306.3$-$0.9.

In this paper, we investigate the X-ray spectral properties and the explosion type of G306.3$-$0.9 based on a $\sim$190 ks {\it Suzaku} observation. We also analyzed 6-years of {\it Fermi}-LAT data and interpreted the results with the help of multi-wavelength data. This paper is organized as follows: The X-ray and gamma-ray spectral analyses are described in Section 2. The results are discussed in Section 3. Summary and Conclusions are given in Section 4.

\section{Analysis}

\subsection{X-ray Observation and Data Reduction}
We used an archival {\it Suzaku} data (Obs ID. 509072010). The observation was performed by the X-ray Imaging Spectrometer \citep[XIS;][]{Ko07} on 2014 August 20 for an exposure time of $\sim$190 ks. The XIS instrument consists of four X-ray CCD cameras on the focal planes of the X-Ray Telescope \citep[XRT;][]{Se07}. The XIS0, 2 and 3 cameras have front-illuminated (FI) CCDs, whereas the XIS1 is back-illuminated (BI). XIS0, XIS1, and XIS3 were available in this observation.  

Data reduction and analysis were made with HEADAS software version 6.16 and {\sc xspec} version 12.9.0 \citep {Ar96} with AtomDB 3.0.3 \citep{Sm01, Fo12}. The 5$\times$5 and 3$\times$3 editing mode event files were combined using \texttt{xis5$\times$5to3$\times$3} and {\sc xselect} version 2.4b. The redistribution matrix file (RMF) and the ancillary response file (ARF) for the spectral analyses are generated by the {\sc xisrmfgen} and {\sc xissimarfgen} tools, respectively \citep {Is07}.

\subsection{X-ray Spectral Analysis} \label{Spectral analysis}
\begin{figure}
\centering \vspace*{1pt}
\includegraphics[width=0.35\textwidth]{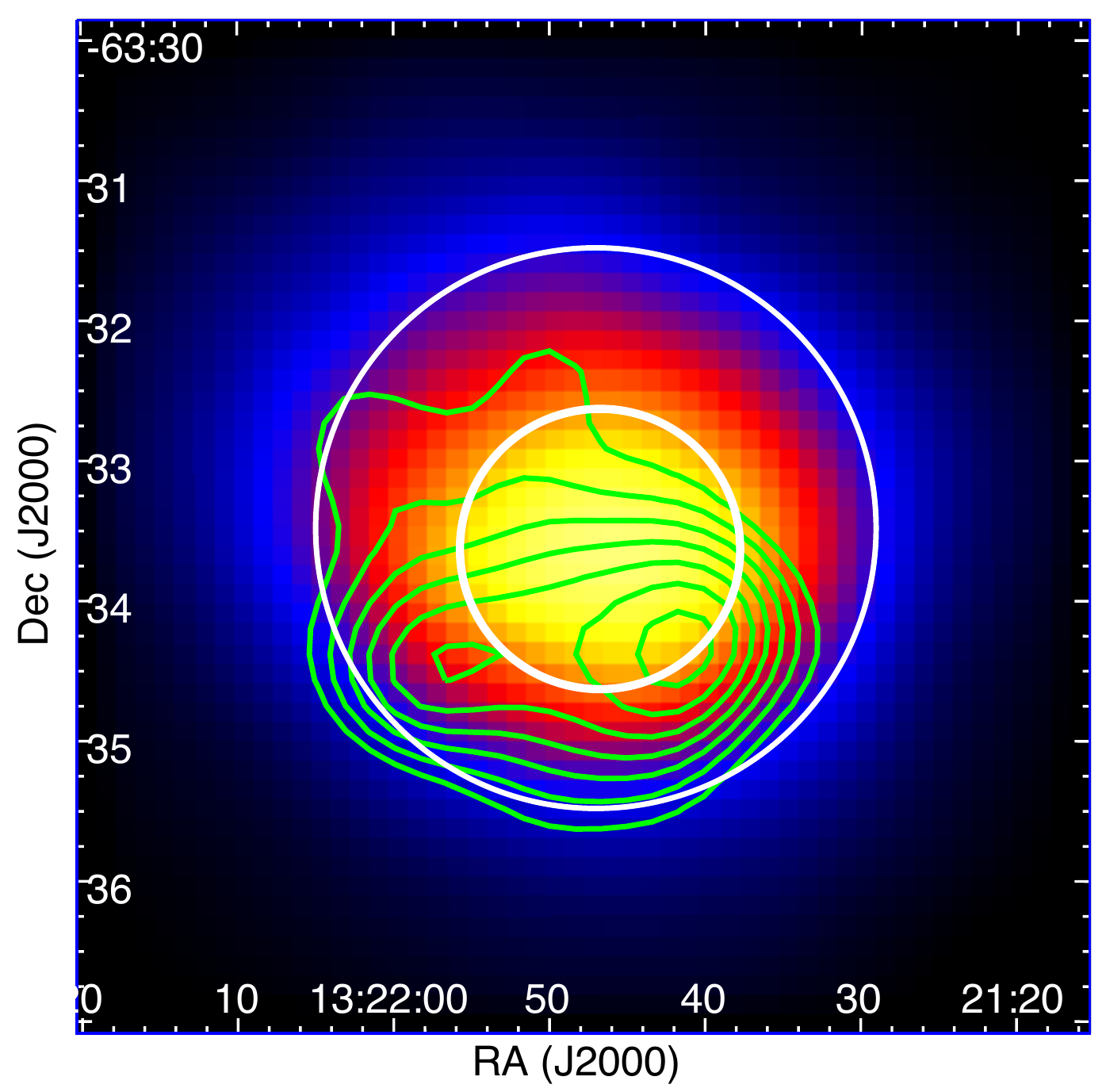}
\includegraphics[width=0.35\textwidth]{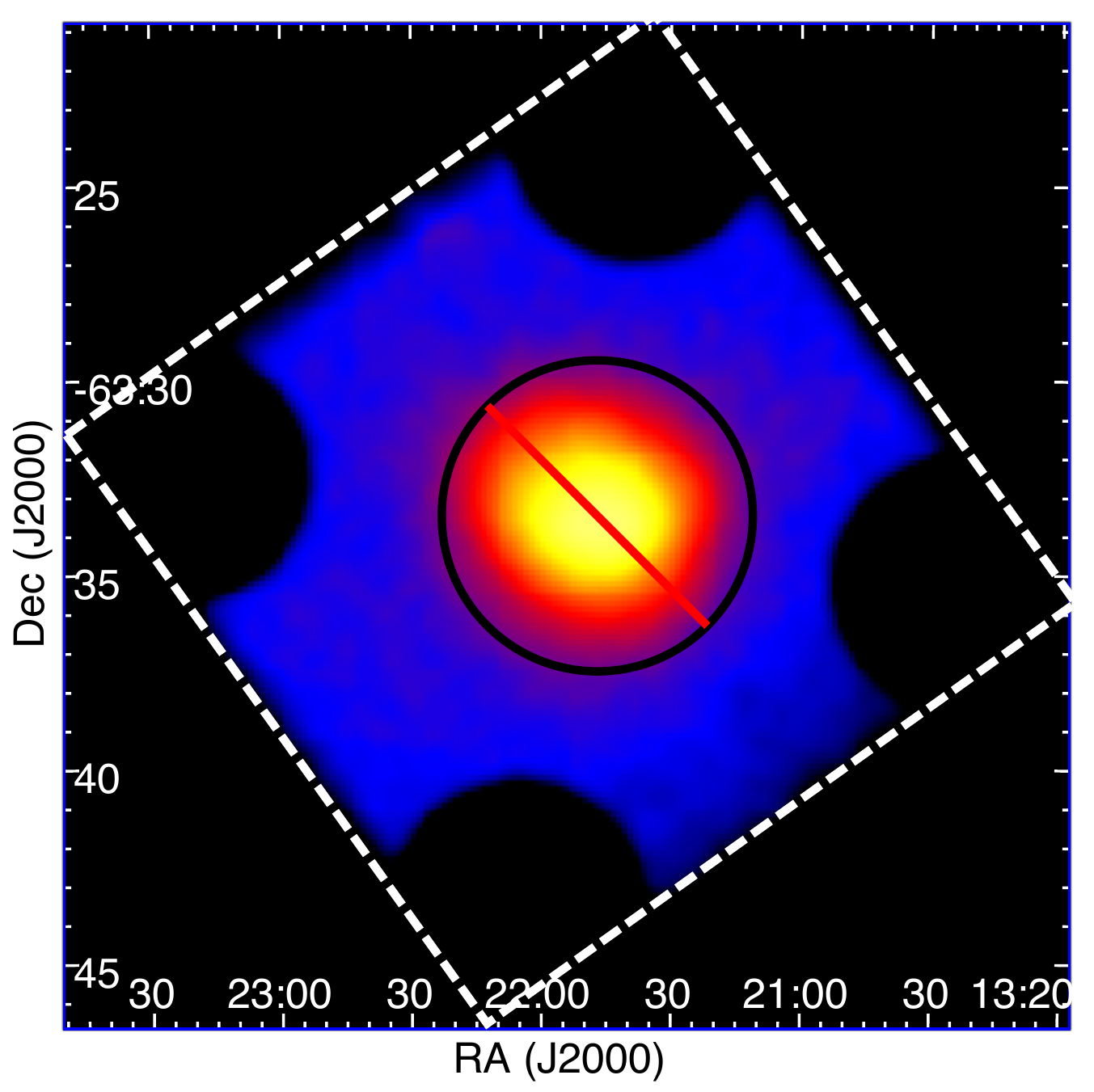}
\caption{Upper panel: {\it Suzaku} XIS image of G306.3$-$0.9 in the 0.3$-$10.0 keV energy band, overlaid with the Molonglo Observatory Synthesis Telescope (MOST) radio contours at 843 MHz \citep {Wh96}. The radio contour levels are 3, 5.2, 7.4, 9.6, 11.8, and 14 mJy beam$^{-1}$. The source regions for the spectra of the whole and the centre are shown by the solid circles. Lower panel: The FoV region of the XIS is shown by the dashed area. For the background estimation, we excluded the calibration regions and the source region from the dashed area.}
\label{figure_1}
\end{figure}

Figure \ref{figure_1} shows XIS image of G306.3$-$0.9 in the 0.3$-$10.0 keV energy band. In order to characterize the emission, we extracted spectra from two circular regions centered at the source location with radii of 1.0 arcmin (central region) and 2.5 arcmin (the whole SNR), and from an annulus region (the rim region). These regions are shown by the solid circles in the upper panel of Figure 1. For comparison, the
radio data (J.A. Combi, private communication) at 843 MHz taken from the MOST Supernova Remnant Catalog \citep {Wh96} are overlaid. The background spectra were extracted from a source free region in the same field of view (FoV), which is shown as the dashed area in the lower panel of Figure 1 excluding the calibration regions.

\subsubsection {Background Estimation}
For G306.3$-$0.9, the background emission contains the non-X-ray background (NXB), cosmic X-ray background (CXB) and Galactic ridge X-ray emission (GRXE). The NXB for the source and the background spectra were extracted from the night-earth data using {\sc xisnxbgen} \citep {Ta08}. The NXB was subtracted from the source and the background data. We fitted the NXB-subtracted background spectrum with a model of Abs1$\times$power-law + Abs2$\times$(apec+apec), where the apec is a collisional ionization equilibrium (CIE) plasma model in the {\sc xspec}. In this fitting, an absorbed two-temperature apec component represents the GRXE emission, while an absorbed power-law model represents the CXB emission. To define the CXB emission, we assumed a power-law shape with a photon index of 1.4, and a surface brightness of $5.4\times10^{-15}$ erg s$^{-1}$ cm$^{-2}$ arcmin$^{-2}$ in the 2$-$10 keV band \citep {Ku02}. Next, we simulated the background spectra using the \texttt{fakeit} command in {\sc xspec} and subtracted it from the source spectra. All spectra were grouped with a minimum of 30 counts bin$^{-1}$.

\subsubsection {Spectral Fitting}

\begin{figure}
\centering \vspace*{1pt}
\includegraphics[width=0.5\textwidth]{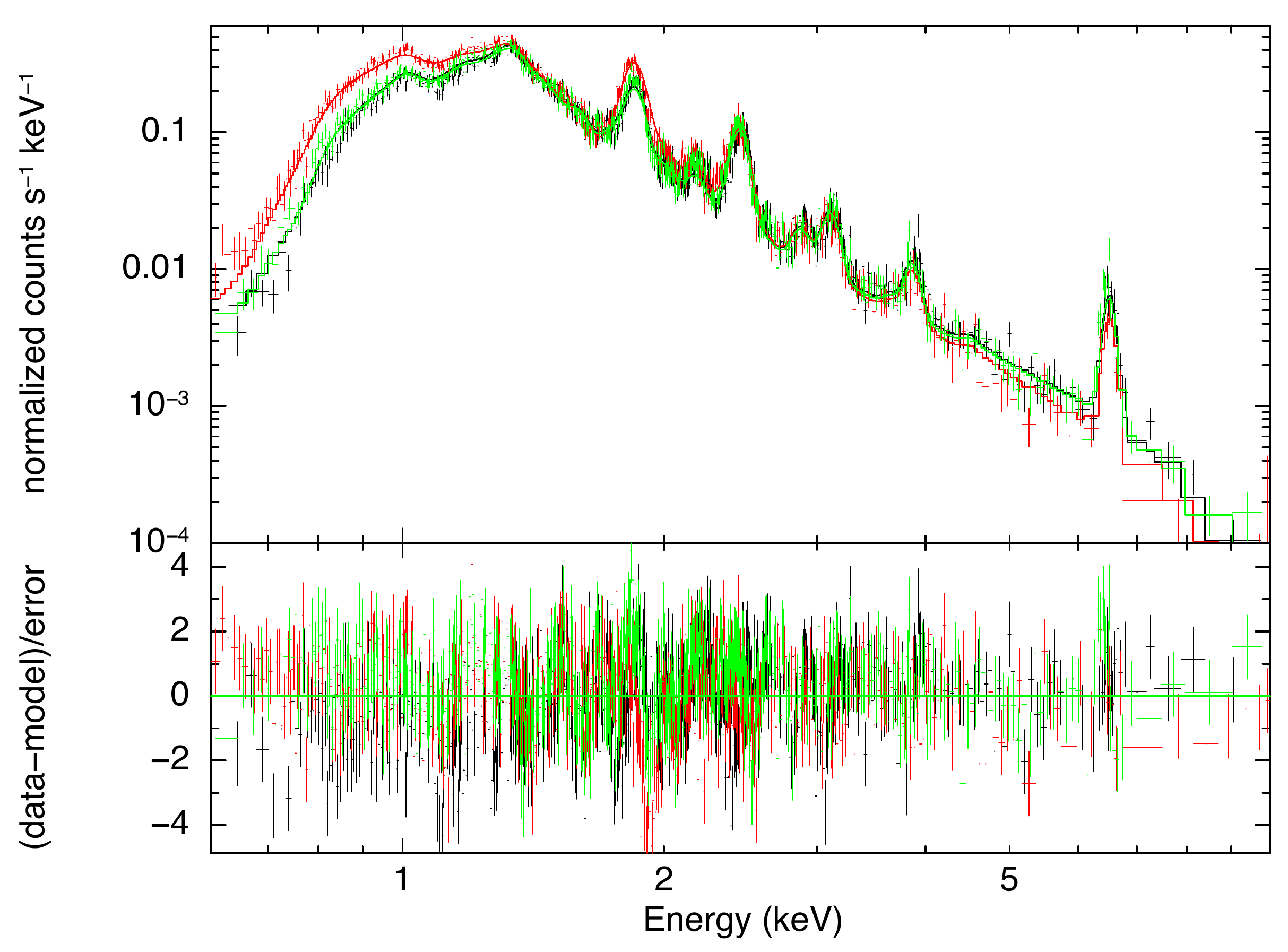}
\caption{Background-subtracted XIS (XIS0, XIS1 and XIS3) spectra of the whole region of G306.3$-$0.9 in the 0.6$-$10.0 keV energy band.  The spectra are overlaid with the best-fit model. The bottom panel is the residuals of the data off the best-fit model.}
\label{figure_2}
\end{figure}

\begin{table*}
 \begin{minipage}{170mm}
  \caption{Best-fitting spectral parameters of G306.3$-$0.9. Abundances are given relative to the solar values of \citet {Wi00}. Errors are at the 90 per cent confidence level.}
\renewcommand{\arraystretch}{1.5}
 \begin{tabular}{@{}p{2.2cm}p{6cm}p{2.3cm}p{2.3cm}p{2.3cm}@{}}
  \hline\hline
     Component& Parameters & Whole & Centre & Rim\\
\hline
Absorption & $N_{\rm H}$ ($\times10^{22}$ cm$^{-2})$ & $1.67_{-0.12}^{+0.25}$  & $1.73_{-0.11}^{+0.29}$      & $1.69_{-0.17}^{+0.34}$ \\

VAPEC & $kT_{\rm e}$ (keV)& $0.59_{-0.01}^{+0.02}$  & $0.61_{-0.02}^{+0.03}$    & $0.49_{-0.01}^{+0.01}$  \\

&  norm ($\times10^{-2}$ ph cm$^{-2}$ s$^{-1}$)  &   $1.83_{-0.24}^{+0.17}$  & $0.89_{-0.07}^{+0.05}$  & $1.45_{-0.24}^{+0.38}$ \\

VNEI & $kT_{\rm e}$ (keV)& $3.17_{-0.31}^{+0.15}$  & $2.64_{-0.41}^{+0.26}$    & $3.06_{-0.33}^{+0.48}$  \\

&   Si  & $6.4_{-0.9}^{+1.1}$  & $3.8_{-0.7}^{+0.6}$    & $2.4_{-0.8}^{+0.9}$  \\

&  S  & $6.6_{-1.3}^{+0.8}$  & $5.9_{-2.2}^{+1.6}$     & $4.4_{-0.9}^{+1.2}$  \\

&  Ar  & $9.7_{-1.9}^{+1.2}$  & $8.9_{-1.1}^{+2.1}$      & $4.9_{-1.3}^{+0.7}$ \\

&  Ca  & $16.6_{-2.4}^{+1.1}$  & $11.3_{-1.8}^{+2.3}$     & $5.1_{-1.5}^{+0.8}$  \\

&  Fe  & $9.9_{-1.3}^{+1.8}$  & $5.7_{-0.6}^{+1.3}$     & $6.2_{-1.1}^{+1.4}$  \\

&   $\tau$ ($\times10^{10}$ cm$^{-3}$ s)& $2.6_{-0.4}^{+0.3}$  & $1.5_{-0.7}^{+0.6}$    & $1.9_{-0.2}^{+0.3}$ \\

&  norm ($\times10^{-4}$ ph cm$^{-2}$ s$^{-1}$)  &   $9.69_{-1.05}^{+1.31}$  & $4.74_{-0.54}^{+0.75}$  & $3.81_{-0.21}^{+1.32}$ \\

\hline

Line      &  Fe K$\alpha$  Energy Centroid (ev)  &    $6504_{-12}^{+10}$       &    $6501_{-17}^{+11}$       &    $6503_{-15}^{+9}$        \\

\hline

& reduced-$\chi^{2}$ (dof) & 1.18 (1966)  &  1.25 (581)  &  (1.21) 461   \\

 \hline
\end{tabular}
\end{minipage}
\end{table*}

We first fit the spectra of the whole region with an absorbed \citep[TBABS:][]{Wi00} single-component variable-abundance non-equilibrium ionization (NEI) plasma model (VNEI model with NEI version 3.0 in {\sc xspec}). In this fitting the absorbtion ($N_{\rm H}$), electron temperature ($kT_{\rm e}$), and ionization parameter ($\tau$=$n_{\rm e}t$) are free parameters, where $n_{\rm e}$ and $t$ are the electron density and elapsed time following shock-heating. The abundances of Si, S, Ar, Ca, and Fe are free parameters, while the other abundances were fixed to the solar abundance \citep {Wi00}. This model gave a large reduced ${\chi}^2$ of 3.28 (dof=1975) with large residuals $\sim$1.20$-$1.23 keV and $\sim$6.5 keV. We refit the spectra with Ne and Mg abundances varied. The fit improved but it was still not statistically acceptable (reduced ${\chi}^2$=2.8). Because the one-component plasma model failed to reproduce the Fe-K$\alpha$ line ($\sim$6.5 keV) profile and the fit gave a large reduced chi-squared value, we fitted the spectra with a two-component thermal plasma model, in CIE (VAPEC model in {\sc xspec}) and VNEI model. The abundances of Si, S, Ar, Ca and Fe for the VNEI component are free parameters. We also allowed the abundances of Ne and Mg to vary freely. We found the abundances of Ne and Mg are near solar values. Therefore, we fixed them to the solar value. For CIE component, the electron temperature and normalization are free parameters, while the abundances of all elements are fixed to the solar values assuming that the emission is from the shocked ISM. This fit leaves a large residual $\sim$1.2 keV that comes from the uncertainty of the Fe L-shell data in the VNEI code. Thus we also added a narrow Gaussian emission line at $\sim$1.2 keV. This additional Gaussian line component does not affect the best-fit values for all other model parameters. The fit was improved to be ${\chi}^2$ = 1.18 (dof=1966). We also fitted the spectra extracted from the centre and the rim regions with an absorbed VAPEC and VNEI model. The best-fitting parameters are summarized in Table 1, and the background-subtracted XIS spectra are shown in Figure \ref{figure_2}.  

\subsection {Gamma-ray Analysis}
To search for a gamma-ray emission in the GeV energy range, we analyzed the gamma-ray data of {\it Fermi}-LAT for the time period of 2008-08-04 $-$ 2016-03-23. In this paper we used the Fermi analysis toolkit \texttt{fermipy}\footnote{http://fermipy.readthedocs.io/en/latest/index.html}.   

Using \texttt{gtselect} of Fermi Science Tools (FST), we selected the {\it Fermi}-LAT Pass 8 `Source' class and front$+$back type events coming from zenith angles smaller than 90$^{\circ}$ and from a circular region of interest (ROI) with a radius of 20$^{\circ}$ centred at the SNR's radio position. The maximum likelihood fitting method \citep{Ma96} was employed on the spatially and spectrally binned data using the P8R2$_{-}$SOURCE$_{-}\!\!$V6 version of the instrument response function. After the maximum likelihood fitting between 200 MeV and 300 GeV, the detection significance value is calculated, which is roughly the square root of the test statistics (TS) value and larger TS values indicate that the null hypothesis (maximum likelihood value for a model without an additional source) is incorrect.

The model of the analysis region contains the diffuse background sources and all the point-like and extended sources from the 3rd {\it Fermi}-LAT Source Catalog \citep{Ac15} located within a square region with side 15$^{\circ}$ centred on the ROI centre. All parameters of the diffuse Galactic emission (\emph{gll$_{-}$iem$_{-}$v6.fits}) and the isotropic component (\emph{iso$_{-}$P8R2$_{-}$SOURCE$_{-}\!\!$V6$_{-}\!$v06.txt}) were freed. We also freed all sources with TS $>$ 10 and fixed all sources with TS $<$ 10.

The TS map created using this model showed gamma-ray excess extending toward the south-west of G306.3$-$0.9. To understand if this excess was related to the SNR, we added this SNR as a point-like source with a PL-type spectral shape into the background model, since there was no gamma-ray source corresponding to G306.3$-$0.9 in the 3rd {\it Fermi}-LAT Source Catalog \citep{Ac15}. After creating a new TS map, including where G306.3$-$0.9 in the background model, south-west of the SNR position still showed a significant amount of gamma-ray excess spread across a wide area. To account for this extended gamma-ray excess, we added a new point-like source with a PL-type spectral shape into the model, which we called `SourceA' and found its best-fitting position. However, introducing SourceA into the background model could not remove the excess of gamma rays in the  south-west of G306.3$-$0.9, which were distributed in a region encircled by the significance contours of 5$\sigma$. Finally, we tested different models of extension (Radial Gaussian and Disk) for SourceA having a PL-type spectrum. The results of these analyses are summarized in Section \ref{subsection:Gamma-rayResults}. 

\section{Results and Discussion}

\subsection {SNR Origin}
Previous X-ray studies of G306.3$-$0.9 suggested a Type Ia progenitor for this remnant \citep {Re13, Co16}. Using {\it Suzaku} XIS data, we investigated the explosive origin of G306.3$-$0.9. For this investigation, we consider the abundance pattern and the centroid of the Fe-K line of the remnant. To compare our data with the SN explosion models, we calculated the abundance ratios of S, Ar, Ca, and Fe relative to Si. Table 2 shows a comparison of our best-fitting relative abundances with the results from the CC models \citep {Wo95} for various progenitor masses and Type Ia models \citep {No97,Ba03}. We also give the abundance ratios of S/Si, Ar/Si and Ca/Si from {\it Chandra}/{\it XMM-Newton} data \citep {Co16} in Table 2. 

The abundance ratios of S/Si, Ar/Si and Ca/Si of {\it Suzaku} and {\it Chandra}/{\it XMM-Newton} data are consistent with 12$M_{\sun}$ CC model as seen Table 2. But our abundance ratio of Fe/Si is significantly higher than that of the 12$M_{\sun}$ CC model. This ratio is consistent with the CC model with the progenitor mass of 11$M_{\sun}$ and W7 model. All four ratios are consistent with one or more of the four Type Ia models. However, none of Type Ia models agrees with more than two ratios. None of the ratios is consistent with the 15$M_{\sun}$ CC model. Therefore, by looking at the results on Table 2, neither Type Ia nor CC SN models are conclusive for G306.3$-$0.9. However, the CC SN of a 15$M_{\sun}$ progenitor is ruled out for the origin of G306.3$-$0.9.

\begin{table*}
\caption{Comparisons of Abundance Ratios between
G306.3$-$0.9 and models.}
\begin{center}
 \begin{minipage}{140mm}
\renewcommand{\arraystretch}{1.5}
 \begin{tabular}{@{}ccccccp{1.2cm}ccp{1.2cm}@{}}
  \hline
    \hline
    &&&\multicolumn{4}{c}{Type Ia Models\footnote{W7 and WDD2 models by \citet {No97}; PDDe and DDTe models by \citet {Ba03}.}}&\multicolumn{3}{c}{CC Models\footnote{\citet {Wo95}.}}\\
    \cline{4-7}
     \cline{8-10}
      Abundance Ratio& {\it Suzaku}\footnote{For the Whole region in our work.} & {\it Chandra}/{\it XMM-Newton}\footnote{\citet {Co16}.} &W7
      & WDD2& PDDe
      & DDTe&11$M_{\sun}$&12$M_{\sun}$&15$M_{\sun}$\\
\hline
S/Si& $1.03_{-0.24}^{+0.22}$  & 1.75 &  1.07   & 1.17   & 1.5  & 1.4  & 0.87 & 1.53 & 0.62\\
Ar/Si  & $1.52_{-0.36}^{+0.32}$ & 1.27   &  0.89   & 1.38   & 0.68 & 0.60 & 0.63 & 1.62 & 0.50\\
Ca/Si  & $2.59_{-0.52}^{+0.48}$ &2.72   &  0.75   & 0.94   & 2.9  & 2.5  & 0.65 & 2.04 & 0.43\\
Fe/Si  & $1.55_{-0.30}^{+0.39}$   & &  1.56   & 0.85   & 0.89 & 0.91 & 1.37 & 0.23 & 0.70\\
  \hline
\end{tabular}
 \end{minipage}
\end{center}
\end{table*}

Recently, \citet {Ya14} systematically searched for Fe-K emission from Galactic and LMC SNRs using {\it Suzaku} data. They concluded that the centroid energy of the Fe-K emission and the ionization state of Fe are a powerful tool for distinguishing progenitor types. They found that Fe-K$\alpha$ centroid energies are below $\sim$6.55 keV for Type Ia SNRs and the Fe-K emission of Type Ia SNRs is significantly less ionized than in CC-SNRs. \citet {Co16} obtained the centroid of the Fe-K line in the central region of the SNR and estimated a centroid of 6.52$\pm$0.01 keV. They concluded that this value consistent with a Type Ia origin. In order to estimate the centroid energy of an Fe-K$\alpha$, we fitted the 5.0$-$8.0 keV spectra with a PL and a Gaussian. We estimated the centroid energy of Fe-K$\alpha$ for each region and listed them in Table 1. The centroid energy of Fe-K$\alpha$ is supportive of the Type Ia SN origin.

\subsection{Gamma-ray Results} \label{subsection:Gamma-rayResults}
We found no excess gamma-ray emission from the direction of G306.3$-$0.9, where the upper limit at 95 per cent confidence level (CL) on the photon flux and energy flux was found to be 1.3 $\times$ 10$^{-8}$ photons cm$^{-2}$ s$^{-1}$ and 5.3 $\times$ 10$^{-6}$ MeV cm$^{-2}$ s$^{-1}$, respectively. 

\begin{figure*}
\centering \vspace*{1pt}
\includegraphics[width=0.8\textwidth]{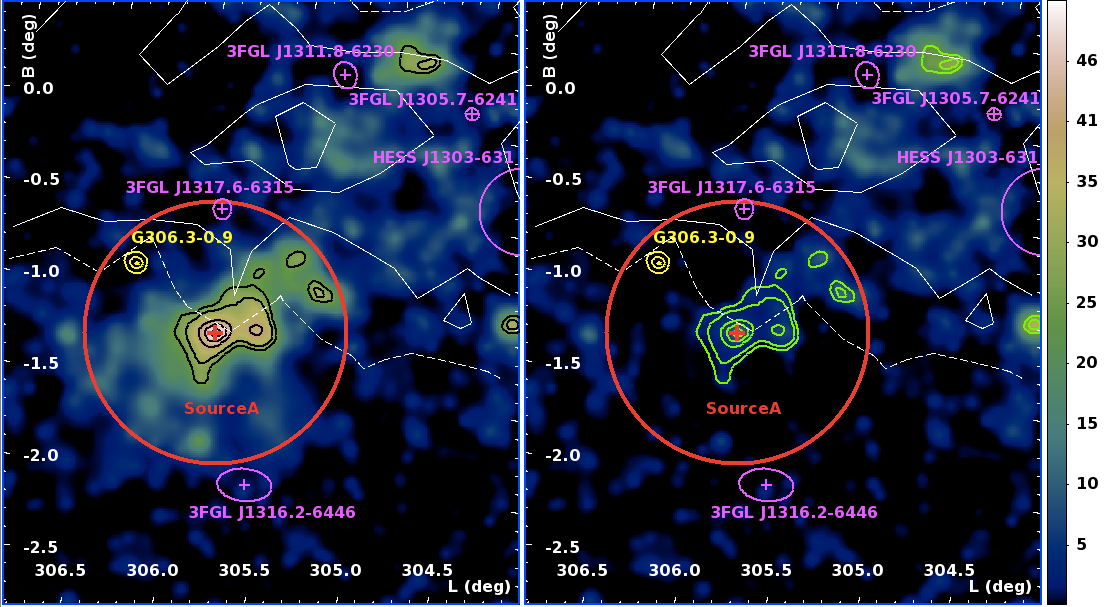}
\caption{The gamma-ray TS map of the neighborhood of G306.3$-$0.9 and SourceA. Left Panel: SourceA is not included in the background model and G306.3$-$0.9 is left in the model as a point-like source. On both panels, {\it Suzaku} X-ray counts (50, 200, 500) are shown in yellow color, white lines show the CO intensity contours of 12.5 (dashed), 30 (solid), 55 (solid), 78 (solid), 101 (solid) K km s$^{-1}$, and the red circle and red cross represent the extension and best-fit location of SourceA, respectively. Fermi-LAT sources are shown in magenta color. Right Panel: Both G306.3$-$0.9 and SourceA are included in the background model. The black contours on the left panel and green contours on the right panel are for the gamma-ray TS values (25, 30, 40, 45, 49).}
\label{gfigure_2}
\end{figure*}

A new gamma-ray source was detected at the south-west of the G306.3$-$0.9 position, which we designate as `SourceA' in this analysis. Assuming that it is a point-like source, the TS value of $\sim$94 was found and the best-fitting location of this source was found as R.A.(J2000) = 199$^{\circ}\!\!$.47 $\pm$ 0$^{\circ}\!\!$.07$^{\rm stat}$ and decl.(J2000) = $-$63$^{\circ}\!\!$.93 $\pm$ 0$^{\circ}\!\!$.07$^{\rm stat}$ using the \texttt{localize} method of FST. This new source has an offset from the radio location of the SNR by an amount of 0$^{\circ}\!\!$.575. At the best-fitting location, we found the spectral index to be $\Gamma$ = 2.7 $\pm$ 0.1 for the PL-type spectrum. The total photon flux and energy flux of SourceA is (1.43 $\pm$ 0.27) $\times$ 10$^{-8}$ photons cm$^{-2}$ s$^{-1}$ and (8.10 $\pm$ 1.08) $\times$ 10$^{-6}$ MeV cm$^{-2}$ s$^{-1}$, respectively, for the point-like source model with PL-type spectrum. 

\begin{figure}
\centering \vspace*{1pt}
\includegraphics[width=0.5\textwidth]{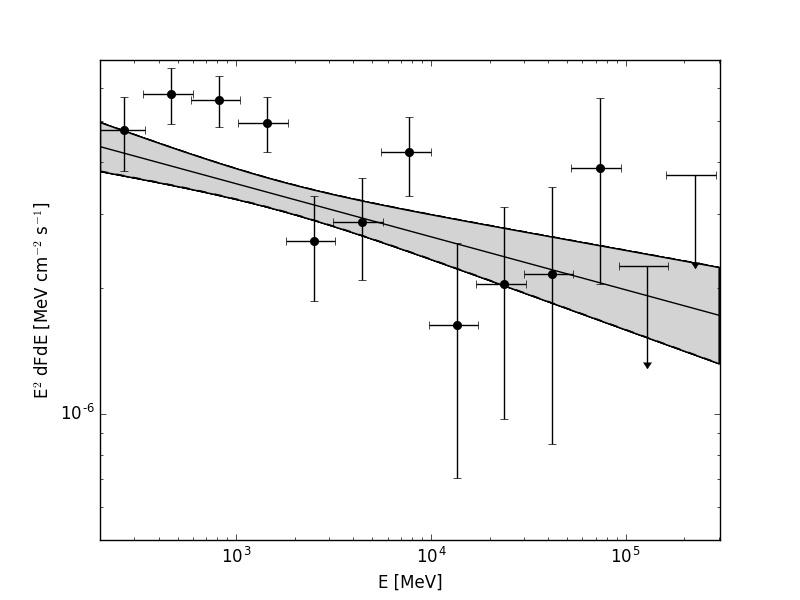}
\caption{The SED of G306.3$-$0.9 created assuming the SNR as a Disk like extended source with a PL-type spectra. The shaded region represents the model flux and its statistical errors obtained from fitting a PL-type spectrum to the given spectral data.}
\label{gfigure_3}
\end{figure}

We used two extension models for the gamma-ray emission morphology of SourceA: Disk and Radial Gaussian models, where the centres of these extension models were kept at the best-fitting location of SourceA. To detect the extension of a source, we used the TS of the extension (TS$_{\rm ext}$) parameter, which is the likelihood ratio comparing the likelihood for being a point-like source (L$_{\rm pt}$) to a likelihood for an existing extension (L$_{\rm ext}$), TS$_{\rm ext}$ = 2log(L$_{\rm ext}$/L$_{\rm pt}$). We tabulated the `Extension Width', which is the 68\% containment radius of the extension model (R$_{68}$), and the corresponding TS$_{\rm ext}$ values of these fits in Table 3. The highest TS$_{\rm ext}$ value was found as $\sim$40 and the total TS value of SourceA was found to be 158 assuming the Disk like extension model. Assuming a PL-type spectrum for this extended source, we obtained $\Gamma$ = 2.1 and the total photon flux and energy flux of SourceA was found to be (1.9 $\pm$ 0.2) $\times$ 10$^{-8}$ photons cm$^{-2}$ s$^{-1}$ and (2.07 $\pm$ 0.2) $\times$ 10$^{-5}$ MeV cm$^{-2}$ s$^{-1}$, respectively. These results found for SourceA are in agreement with the ones that were given for G306.3$-$0.8 in the 1st {\it Fermi}-LAT SNR Catalog \citep{Ac16}. The TS map shown on the Right Panel of Figure \ref{gfigure_2} was obtained after applying the Disk like extension model to SourceA and adding SourceA to the background model. The spectral energy distribution (SED) of SourceA with a Disk extension and PL-type spectrum is shown in Figure \ref{gfigure_3}.

The upper limit at 95 per cent CL on the energy flux of G306.3$-$0.9 was found to be 2.7 $\times$ 10$^{-6}$ MeV cm$^{-2}$ s$^{-1}$ adding SourceA as a point-like source with a PL-type spectrum and 3.1 $\times$ 10$^{-6}$ MeV cm$^{-2}$ s$^{-1}$ adding SourceA as an extended source with a PL-type spectrum. 

\begin{table}
 \begin{minipage}{90mm}
 \begin{center}
  \caption{Gamma-ray extension model fits for G306.3$-$0.9.}
 \begin{tabular}{@{}ccc@{}}
  \hline\hline
       Model & Extension Width (R$_{68}$)  & TS$_{\rm ext}$\\
\hline
Radial Gaussian & 0$^{\circ}\!\!$.4171 $+$ 0$^{\circ}\!\!$.0693 $-$ 0$^{\circ}\!\!$.0823 & 25.1 \\
Disk                    & 0$^{\circ}\!\!$.7256 $+$ 0$^{\circ}\!\!$.0713 $-$ 0$^{\circ}\!\!$.0713 & 39.8 \\
 \hline
\end{tabular}
\end{center}
\end{minipage}
\end{table}

\subsection {Gamma-ray Variability and Pulsation}
\begin{figure}
\centering \vspace*{1pt}
\includegraphics[width=0.5\textwidth]{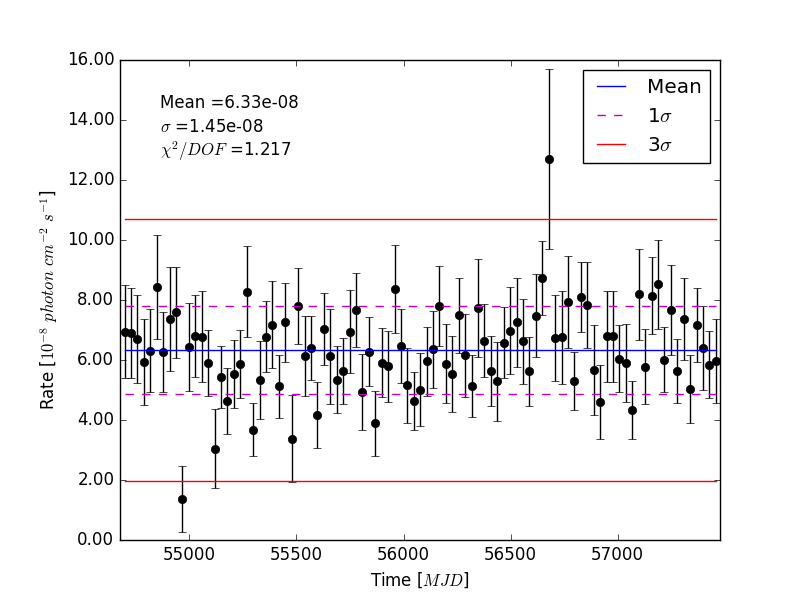}
\caption{The monthly gamma-ray variability for SourceA assuming a PL type spectrum in the energy range of 0.2 $-$ 300 GeV.}
\label{gfigure_4}
\end{figure}

To see the long term variability in the light curve of SourceA, we apply {\it Fermi}-LAT {\it aperture photometry} taking data from the circular region of 0.2$^{\circ}$ around the best-fitting position of SourceA. Figure \ref{gfigure_4} shows the 1-month binned light curve, where we checked for possible variations in the flux levels. In Figure \ref{gfigure_4} most of the flux data points remain within the 1$\sigma$ and 3$\sigma$ bands. One of the flux data points with large error bars is above 3$\sigma$, could be due to the contamination by the nearby pulsar 3FGL J1317.6$-$6315. Therefore, by looking at 0.$\!\!^{\circ}$2 around the best-fitting location of SourceA we conclude that SourceA shows no variability including pulsations in gamma rays.

\subsection {The Molecular Environment}
To investigate the molecular environment around G306.3$-$0.9 and SourceA, we used the carbon monoxide (CO) data collected by the Harvard-Smithsonian CfA 1.2 m Millimeter-Wave Telescope from the fourth quadrant (DHT36)\footnote{https://www.cfa.harvard.edu/rtdc/CO/NumberedRegions/DHT36/index.html} with a 0.$\!\!^{\circ}$25 beam sampling \citep{Br89}.

\begin{figure*}
\centering
\includegraphics[width=0.8\textwidth]{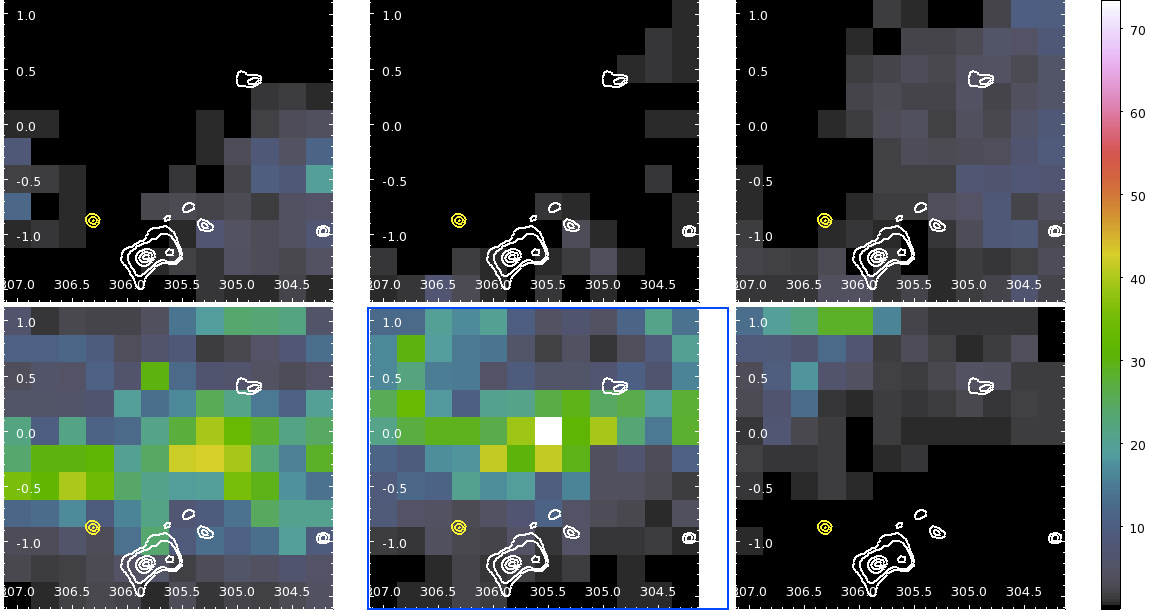}
\caption{CO intensity maps produced at different velocity ranges of [$-$70,$-$50], [$-$50,$-$35], [$-$35,$-$10] km s$^{-1}$ for the panels from bottom right to left and [$-$10,0], [0,15], [15,40] km s$^{-1}$ from top right to left panels. The white contours represent the gamma-ray TS values of SourceA for 25, 30, 40, 45, and 49 and yellow contours are the X-ray counts at 50, 200 and 500. The color scale for the CO intensity is set between 0.0 and 76.2 K km s$^{-1}$ for all panels.}
\label{gfigure_5}
\end{figure*}

We checked the velocity integrated CO intensity (W$_{\rm CO}$) in the whole velocity range integrated from $-$70 to $+$40 km  s$^{-1}$, where the velocity intervals are divided such that each range includes at least one cloud cluster peaking in temperature at a certain velocity.  Figure \ref{gfigure_5} shows the W$_{\rm CO}$ maps produced at different velocity ranges of [$-$70,$-$50], [$-$50,$-$35], [$-$35,$-$10] km s$^{-1}$ from bottom right to left and [$-$10,0], [0,15], [15,40] km s$^{-1}$ from top right to left. The white contours represent the TS values of SourceA gamma-ray data at 25, 30, 40, and 45, and yellow contours are the X-ray counts at 50, 200 and 500. The color scale for W$_{\rm CO}$ is set to the same range for all plots, which is between 0.0 and 76.2 K km s$^{-1}$, with the W$_{\rm CO}$ values peaking in the velocity range of $[-$50,$-$35] km s$^{-1}$. Calculating the distance for this velocity range, we obtain a distance range of $\sim$3$-$6 kpc. The distance of the clouds that coincide with SourceA are in the velocity range of [$-$35,$-$10] km s$^{-1}$, where the distance range calculated is $\sim$1$-$9 kpc. The distance range found here is consistent with other measurements \citep{Re13,Co16}, but it is not much constraining. So, we will use the distance to the SNR as 8 kpc assuming that the SNR lies at a fiducial distance of the Galactic center.

The total W$_{\rm CO}$ value found for the regions overlapping with SourceA and G306.3$-$0.9 is about 44 and 3 K km s$^{-1}$, respectively. Using the CO-to-H$_2$ conversion factor of X = 1.8 $\times$ 10$^{20}$ cm$^{-2}$ K$^{-1}$ km s$^{-1}$ \citep{dame2001}, we found N(H$_2$) = 0.8 $\times$ 10$^{22}$ cm$^{-2}$ for SourceA and N(H$_2$) = 0.5 $\times$ 10$^{21}$ cm$^{-2}$ for G306.3$-$0.9. 

There are two gamma-ray production scenarios involving molecular clouds (MCs): First model is the `interacting cloud model'\citep{Ah94, In12}, where gamma rays are produced through the interaction of accelerated hadrons like protons with the molecular material of the MC producing neutral-pions, which then decay into gamma rays. In the second scenario, the `illuminated cloud scenario'\citep{Yam06, Ga09, Oh11}, cosmic rays (CRs) escape from the SNR and diffuse into the nearby dense MCs producing gamma rays through the neutral-pion decay. 

The best evidence for the interacting cloud scenario, is to detect maser emission at the location of the SNR interacting with MCs. There are no masers reported at the locations of G306.3$-$0.9 and SourceA indicating an interaction with the MCs. By assuming a spherical geometry of the cloud, we computed the average density of protons to be $\sim$3 protons cm$^{-3}$ $\sim$14 protons cm$^{-3}$ for G306.3$-$0.9 and SourceA, respectively. These derived proton densities are much lower in comparison to the densities derived for other SNRs, such as 3C 391 and G349.7$+$0.2 \citep{Er14, Er15}, that are interacting with MCs.

However, although CO is the most widely used proxy to track down molecular gas in the Galaxy, not all portions of an MC containing H$_2$ also contain CO. New data from ESA's Herschel Space Observatory ({\it Herschel}) \citep{Pi13} confirmed this estimation by showing that almost one third of all molecular gas in the Milky Way had remained undetected. {\it Herschel}'s survey has established the three-dimensional distribution of the molecular gas across the Milky Way using a different tracer (ionized carbon (C+)) for H$_2$. The densest MC cores of the ISM, where most of the molecular gas resides, contain both H$_2$ and CO, but their immediate surroundings might be influenced by ultraviolet radiation from nearby stars, where H$_2$ and C+ are found to coexist.  `CO-dark' molecular gas is named for regions of H$_2$ mixed with CO that can not be traced by CO emission. The {\it Herschel} data showed that CO-dark H$_2$ accounts for about 30 per cent of the Milky Way's entire reservoir of molecular gas \citep{Pi13}. 

Assuming there is a dense gas cloud at the location of SourceA, we can interpret the results such that the CRs escaping from G306.3$-$0.9 reaching the dense MC at the location of SourceA, causing the emission of gamma rays (the illuminated cloud scenario). SourceA has a radius of extension of 0.$\!\!^{\circ}$72 and the angular distance between the radio location of G306.3$-$0.9 and best-fitting location of SourceA is 0.$\!\!^{\circ}$576, which corresponds to a physical distance of 81 pc considering an 8 kpc distance to the SNR and assuming that SourceA is also at the same distance as the SNR. \citet{Ga09} has shown in their Figure 5 left-most panel, the gamma-ray energy flux emitted from an MC of mass 10$^5$ solar masses and at a distance of 1 kpc for the case, where the distance between the SNR and MC is about 50 pc. 
Using the dotted line, which refers to the emission at a time 2000 years after the SN explosion, we calculated the expected gamma-ray energy flux emitted for 1 kpc of distance, which is about 3 $\times$ 10$^{-6}$ MeV cm$^{-2}$ s$^{-1}$ and 3 $\times$ 10$^{-7}$ MeV cm$^{-2}$ s$^{-1}$ for 10 and 100 GeV, respectively. After correcting for the distance of 8 kpc, these values become $\sim$4 $\times$ 10$^{-8}$ MeV cm$^{-2}$ s$^{-1}$ and $\sim$4 $\times$ 10$^{-9}$ MeV cm$^{-2}$ s$^{-1}$ for 10 and 100 GeV, respectively. If the distance to SourceA is about 1 kpc, the SED data point at 10 GeV of SourceA shown in Figure \ref{gfigure_3} is comparable to the expected gamma-ray energy flux from the illuminated clouds. If the distance of SourceA is the assumed 8 kpc, comparing the estimated gamma-ray energy flux values of the illuminated clouds with the measured SED data points of SourceA shown in Figure \ref{gfigure_3}, we can conclude that the estimated gamma-ray energy flux values are lower than the measured ones. So, there might be a source of gamma rays other than the illuminated clouds that is contributing to the total gamma-ray emission coming from SourceA.

\section{Summary and Conclusions}
In this work, we examined the elemental abundances, ionization state of plasma and the explosive origin of G306.3$-$0.9 using {\it Suzaku} observation and investigated the gamma-ray emission from G306.3$-$0.9 and its neighborhood using the {\it Fermi}-LAT data. Our main conclusions can be summarized as follows:

\begin{enumerate}
\item We found that the thermal X-ray emission from G306.3$-$0.9 consists of two $kT_{\rm e}$ plasmas. The overabundances of Si, S, Ar, Ca and Fe in the hot temperature component confirm the ejecta-dominated nature of G306.3$-$0.9. The low-temperature component is associated with an ISM material. The regional spectral analysis shows that $kT_{\rm e}$, abundances, and $\tau$ of the NEI component are generally highest for the Whole spectrum, compared to the mean values found for the Center and the Rim. For example, abundances of Si and Fe are significantly higher for the Whole SNR than those for the Center and the Rim, which could be a result of systematic uncertainties. These systematic errors do not affect the conclusions drawn in this paper.

\item We clearly detected the Fe-K line emission in the ejecta component of this remnant. Its centroid energy is supportive of the Type Ia SN origin.

\item We compared the results of our spectral fit to the predicted abundances from CC \citep {Wo95} and Type Ia SN \citep {No97,Ba03} models. 
The results show that neither Type Ia nor CC SN of 11$M_{\sun}$ and 12$M_{\sun}$ progenitors is conclusively favored for G306.3$-$0.9, while the CC SN of a 15$M_{\sun}$ progenitor is ruled out.

\item G306.3$-$0.9 is not detected in gamma-rays. X-ray observations of G306.3$-$0.9 revealed that the SNR is formed in a supernova explosion of Type Ia and there is no compact object. Additionally, synchrotron emission from a PWN or X-ray filaments in the shell of the SNR was not observed. So, we do not expect to see gamma-ray emission from G306.3$-$0.9 that could be produced by electrons accelerated at the forward shock through the relativistic bremsstrahlung or inverse Compton scattering processes. The alternative way of producing gamma rays would be through the hadronic process, where accelerated hadrons interact with the background gas and subsequently produce gamma rays from the neutral pion decay. If G306.3$-$0.9 were in a dense molecular cloud region, we might have been able to detect hadronic gamma rays.

\item A new extended gamma-ray source was located in the south-west of G306.3$-$0.9, which we called `SourceA' in our paper. The best-fitting location of R.A.(J2000) = 199$^{\circ}\!\!$.47 $\pm$ 0$^{\circ}\!\!$.07$^{\rm stat}$ and decl.(J2000) = $-$63$^{\circ}\!\!$.93 $\pm$ 0$^{\circ}\!\!$.07$^{\rm stat}$ and the extension parameters, as well as the spectral parameters found for SourceA shows that this source is probably `G306.3$-$00.8' reported in the 1st {\it Fermi}-LAT SNR Catalog \citep{Ac16}. 

\item No variations or pulsations were detected in the gamma-ray light curve of SourceA by looking at 0$^{\circ}\!\!$.2 around the best-fitting location of the SNR, eliminating scenarios with variable source types, such as pulsars or binary systems. 

\item SourceA might be an independent source of G306.3$-$0.9. Due to the low molecular gas density at the location of SourceA, a significant contribution of the hadronic gamma-ray emission is not expected. All five pulsar wind nebulae (PWNe) detected by Fermi-LAT (Crab Nebula \citep{Ab10a}, Vela X \citep{Ab10b}, MSH 15$-$52 \citep{Ab10c}, 3C 58 \citep{Ab13b}, HESS J1640$-$465 \citep{Sl10}) have nearly flat spectrum at the GeV energy range. The observed spectrum of SourceA shown in Figure \ref{gfigure_3} is generally consistent with that of a PWN. However, the H.E.S.S. Galactic Plane Survey \citep{Do16} has not reported a detection from the direction of SourceA at a sensitivity level of 1$-$2 per cent of the Crab Nebula, which corresponds to about 6.2 $\times$ 10$^{-6}$ MeV cm$^{-2}$ s$^{-1}$. \citet{Ta13} showed the properties and the TeV detectability of the non-TeV PWNe. SourceA could be a member of this group of PWNe, but there are no observations in other wave-bands, especially in radio and X-rays, at the location of SourceA so far, which  could give some clues on the nature of this mysterious object.

\end{enumerate}

\section*{Acknowledgments}

We thank to Dr. Jorge Ariel Combi for providing us the MOST radio data. We appreciate Dr. Shuta Tanaka's input to the interpretation of the results. Additionally, we thank the referee for his/her constructive comments and recommendations. AS is supported by the Scientific and Technological Research Council of Turkey (T\"{U}B\.{I}TAK) through the B\.{I}DEB-2219 fellowship program. TE thanks to the support by the Young Scientist Award Program (BAGEP-2015). RY is supported in part by grant-in-aid from the Ministry of Education, Culture, Sports, Science, and Technology (MEXT) of Japan, No. 15K05088.

$~$

$~$

{\it Facility}: {\it Suzaku}, {\it Fermi}, {\it Harvard-Smithsonian Center for Astrophysics 1.2 m MMW-radio Telescope, Molonglo Observatory Synthesis Telescope}. 


\onecolumn

\twocolumn


\begin{thebibliography}{99}


\bibitem[\protect\citeauthoryear{Abdo et al.}{2010a}]{Ab10a} Abdo A. A. et al., 2010a, ApJ, 708, 1254

\bibitem[\protect\citeauthoryear{Abdo et al.}{2010b}]{Ab10b} Abdo A. A. et al., 2010b, ApJ, 713, 146

\bibitem[\protect\citeauthoryear{Abdo et al.}{2010c}]{Ab10c} Abdo A. A. et al., 2010c, ApJ, 714, 927

\bibitem[\protect\citeauthoryear{Abdo et al.}{2013a}]{Ab13a} Abdo A. A. et al., 2013a, ApJS, 208, 59 

\bibitem[\protect\citeauthoryear{Abdo et al.}{2013b}]{Ab13b} Abdo A. A. et al. 2013b, ApJS, 208, 17

\bibitem[\protect\citeauthoryear{Acero et al.}{2015}]{Ac15} Acero F. et al., 2015, AJSS, 218, 41 

\bibitem[\protect\citeauthoryear{Acero et al.}{2016}]{Ac16} Acero F. et al., 2016, AJSS, 224, 8 
\bibitem[\protect\citeauthoryear{Aharonian, Drury \& Volk}{1994}]{Ah94}Aharonian F. A., Drury L. O’C., Volk H. J., 1994, A\&A, 285, 645 

\bibitem[\protect\citeauthoryear{Aharonian et al.}{2005}]{Ah05} Aharonian F. et al., 2005, MNRAS, 439, 1013 


\bibitem[\protect\citeauthoryear{Arnaud}{1996}]{Ar96} Arnaud K. A., 1996, in Jacoby G., Barnes J., eds, ASP Conf. Ser. Vol.101, Astronomical Data Analysis Software and Systems V. Astron. Soc. Pac., San Francisco, p. 17

\bibitem[\protect\citeauthoryear{Badenes et al.}{2003}]{Ba03}Badenes C., Bravo E., Borkowski K. J., Dominguez I., 2003, ApJ, 593, 358

\bibitem[\protect\citeauthoryear{Bronfman et al.}{1989}]{Br89} Bronfman L., Alvarez H., Cohen R. S., Thaddeus P., 1989, ApJS, 71, 481

\bibitem[\protect\citeauthoryear{Carey et al.}{2009}]{Ca09}Carey S. J. et al., 2009, PASP, 121, 76

\bibitem[\protect\citeauthoryear{Combi et al.}{2016}]{Co16}Combi J. A., Garcia F., Suarez A. E., Luque-Escamilla P. L., Paron, S. Miceli M., 2016, A\&A, 592A, 125C

\bibitem[\protect\citeauthoryear{Dame, Hartmann \& Thaddeus}{2001}]{dame2001} Dame T. M., Hartmann D., Thaddeus P., 2001, ApJ, 547, 792

\bibitem[\protect\citeauthoryear{Donath et al.}{2016}]{Do16} Donath A. et al., 2016, 6th International Symposium on High-Energy Gamma-Ray Astronomy (Gamma2016), July 11-15, 2016, in Heidelberg, Germany. \footnote{http://www.mpi-hd.mpg.de/hd2016/pages/presentations/Donath.pdf} 

\bibitem[\protect\citeauthoryear{Ergin et al.}{2014}]{Er14} Ergin T., Sezer A., Saha L., Majumdar P., Chatterjee A., Bay{\i}rl{\i} A., Ercan E. N., 2014, ApJ, 790, 65

\bibitem[\protect\citeauthoryear{Ergin et al.}{2015}]{Er15} Ergin T., Sezer A., Saha L., Majumdar P., G\"{o}k F., Ercan E. N., 2015, ApJ, 804, 124

\bibitem[\protect\citeauthoryear{Foster et al.}{2012}]{Fo12} Foster A. R., Ji L., Smith R. K., Brickhouse N. S., 2012, ApJ, 756, 128

\bibitem[\protect\citeauthoryear{Gabici, Aharonian \& Casanova}{2009}]{Ga09} Gabici S., Aharonian F. A., Casanova S., 2009, MNRAS, 396, 1629.

\bibitem[\protect\citeauthoryear{H.E.S.S. Collaboration}{2011}]{HE11} H.E.S.S. Collaboration 2011, https://www.mpi-hd.mpg.de/hfm/HESS/pages/home/som/2011/01/

\bibitem[\protect\citeauthoryear{Inoue et al.}{2012}]{In12}Inoue T., Yamazaki R., Inutsuka S., Fukui Y., 2012, ApJ, 744, 71

\bibitem[\protect\citeauthoryear{Ishisaki et al.}{2007}]{Is07} Ishisaki Y. et al., 2007, PASJ, 59, 113

\bibitem[\protect\citeauthoryear{Koyama et al.}{2007}]{Ko07} Koyama K. et al., 2007, PASJ, 59, 23

\bibitem[\protect\citeauthoryear{Kushino et al.}{2002}]{Ku02}Kushino A., Ishisaki Y., Morita U., Yamasaki N. Y., Ishida M., Ohashi T., Ueda Y., 2002, PASJ, 54, 327

\bibitem[\protect\citeauthoryear{Mattox et al.}{1996}]{Ma96} Mattox, J. R. et al. 1996, ApJ, 461, 396

\bibitem[\protect\citeauthoryear{Nomoto et al.}{1997}]{No97} Nomoto K., Iwamoto K., Nakasato N., Thielemann F.-K., Brachwitz F., Tsujimoto T., Kubo Y., Kishimoto N., 1997, NuPhA, 621, 467

\bibitem[\protect\citeauthoryear{Ohira, Murase \& Yamazaki}{2011}]{Oh11}Ohira Y., Murase K., Yamazaki R., 2011, MNRAS, 410, 1577

\bibitem[\protect\citeauthoryear{Pineda et al.}{2013}]{Pi13} Pineda J. L., et al., 2013, A\&A, 554, A103

\bibitem[\protect\citeauthoryear{Reynolds et al.}{2013}] {Re13} Reynolds M. T. et al., 2013, ApJ, 766, 112

\bibitem[\protect\citeauthoryear{Saz Parkinson et al.}{2016}] {Saz16} Saz Parkinson P. M., Xu H., Yu P. L. H., Salvetti D., Marelli M., Falcone A. D., 2016, ApJ, 820, 20

\bibitem[\protect\citeauthoryear{Serlemitsos et al.}{2007}]{Se07} Serlemitsos P. J. et al., 2007, PASJ, 59, 9

\bibitem[\protect\citeauthoryear{Slane et al.}{2010}]{Sl10} Slane P., Castro D., Funk S., Uchiyama Y., Lemiere A., Gelfand J. D., Lemoine-Goumard M., 2010, ApJ, 720, 266

\bibitem[\protect\citeauthoryear{Smith et al.}{2001}]{Sm01}Smith R. K., Brickhouse N. S., Liedahl D. A., Raymond J. C., 2001, ApJ, 556, L91

\bibitem[\protect\citeauthoryear{Tanaka et al.}{2013}]{Ta13} Tanaka S. J., Takahara F., 2013, MNRAS 429, 2945

\bibitem[\protect\citeauthoryear{Tawa et al.}{2008}]{Ta08} Tawa N. et al., 2008, PASJ, 60, 11

\bibitem[\protect\citeauthoryear{Whiteoak \& Green}{1996}]{Wh96} Whiteoak J. B. Z., Green A. J., 1996, A\&A, 118, 329.

\bibitem[\protect\citeauthoryear{Wilms, Allen \& McCray}{2000}]{Wi00}Wilms J., Allen A., McCray R., 2000, ApJ, 542, 914

\bibitem[\protect\citeauthoryear{Woosley \& Weaver}{1995}]{Wo95} Woosley S. E., Weaver T. A., 1995, ApJS, 101, 181

\bibitem[\protect\citeauthoryear{Wright et al.}{2010}]{Wr10}Wright E. L. et al., 2010, AJ, 140, 1868

\bibitem[\protect\citeauthoryear{Yamaguchi et al.}{2014}]{Ya14}Yamaguchi H. et al., 2014, ApJL, 785, L27

\bibitem[\protect\citeauthoryear{Yamazaki et al.}{2006}]{Yam06}Yamazaki R., Kohri K., Bamba A., Yoshida T., Tsuribe T., Takahara F.,
2006, MNRAS, 371, 1975

\end{thebibliography}
\end{document}